\documentstyle[art12]{article}
\catcode`\@=11 \@addtoreset{equation}{section}\catcode`\@=12
\begin{document}
\begin{center}
{\large
INTEGRATION OF D-DIMENSIONAL 2-FACTOR SPACES\ COSMOLOGICAL
MODELS BY REDUCING TO THE\ GENERALIZED EMDEN-FOWLER EQUATION}
\vskip 5mm
V. R. Gavrilov, V.N. Melnikov\\

{\footnotesize {\it Center for Gravitation and Fundamental Metrology,
     VNIIMS,\\ 3-1 Ulyanova St., Moscow, 117313, Russia}\\
     e-mail: melnikov@fund.phys.msu.su}\\
\vspace{5mm}
{\bf Abstract}
\end{center}
The $D$-dimensional cosmological model on the manifold
$M = R \times M_{1} \times M_{2}$ describing the evolution of 2 Einsteinian
factor spaces, $M_1$ and $M_2$, in the presence of multicomponent perfect fluid
source is considered. The barotropic equation of state for mass-energy
densities and the pressures of the components is assumed in each space.  When
the number of the non Ricci-flat factor spaces and the
number of the perfect fluid components are both equal to 2, the Einstein
equations for the model are reduced to the generalized Emden-Fowler
(second-order ordinary differential) equation, which has been recently
investigated  by Zaitsev and Polyanin within discrete-group analysis.  Using
the integrable classes of this equation one generates the integrable
cosmological models.  The corresponding metrics are presented.  The method
is demonstrated for the special model with Ricci-flat spaces $M_1,M_2$ and
the 2-component perfect fluid source.

PACS numbers: 04.20.J, 04.60.+n, 03.65.Ge

\section{Introduction}
Following the purpose to study the early universe we develop the
multidimensional generalization \cite{G95}-\cite{G97},
\cite{I88}-\cite{I95},\cite{M94},\cite{M95},\cite{R96} of the standard
Friedman-Robertson-Walker world model.
If the extra dimensions of the space-time manifold really exist, the unique
conceivable site, where they might become dynamically important, seems to be
possible. This is some early stage of the evolution. Usually within
multidimensional cosmology (see, for instance, \cite{Bl91}-\cite{Bl94},
\cite{F}-\cite{G97},\cite{Gl}-\cite{M95},\cite{R96},\cite{Sahdev}-\cite{W}
and references therein) it is assumed the occurrence of the topological
partition for the multidimensional space-time on the external 3-dimensional
space and additional so called internal space (or spaces) due to the quantum
processes at the beginning of this stage. In correspondence with such
partition the space-time acquires the topology $M = R \times M_{1} \times
\ldots \times M_{n}$, where $R$ is the time axis, one part of the manifolds
$M_1,\ldots,M_n$ is interpreted as 3-dimensional external space and the
other part stands for internal spaces.  Usually the internal spaces are
compact, however the models with noncompact internal spaces are also
discussed \cite{Gib},\cite{I94}, \cite{RZ}
\cite{Rub}. The subsequent evolution of the
multidimensional Universe is considered as classical admitting the
description by means of the multidimensional Einstein equations. Achieving
the integrability of these equations is the main goal of our investigation.
As the present world seems to be 4-dimensional, there is the assumption that
the internal space(s) had contracted to extremely small sizes, which are
inaccessible for experiment.  This contraction accompanied by the expansion
of the external space is described by some models (the first model of such
type has been found in \cite{C}) within multidimensional cosmology and is
called dynamical compactification.

We consider a mixture of several perfect fluid components as a source for
the multidimensional Einstein equations. Such multicomponent systems are
usually employed in 4-dimensional cosmology and are quite adequate type of
matter for description some early epochs in the history of the universe
\cite{Borner}.

The paper is organized as follows. In section 2 we describe the
multidimensional cosmological model and obtain the Einstein equations in
the form of the Lagrange-Euler equations following from some Lagrangian.
Here we develop the $n$-dimensional vector formalism for the integrating
of the equations of motion. Concluding Section 2 we present the review of
the all known integrable models. In Section 3 we suggest the method for
obtaining the new class of the integrable models on the manifold
$M = R \times M_{1} \times M_{2}$. The method is based on the reducing of
the Einstein equations to the generalized  Emden-Fowler
(second-order ordinary differential) equation. The method is useful for any
2-component model on the manifold $M = R \times M_{1} \times M_{2}$
except for the cases admitting the integration by more simple way.
The total number of the model components is equal to the sum of the number
of the non Ricci-flat spaces  with the number of the perfect
fluid components. The integrable classes recently derived by Zaitsev and
Polyanin of the generalized Emden-Fowler equation allow to generate the new
integrable cosmological models. Their metrics are presented. In Section 3
the method is applied for the models with Ricci-flat spaces $M_{1},M_{2}$
and 2-component perfect fluid.

\section{The model and the equations of motion}

Within $n$-factor spaces  cosmological model
$D$-dimensional space-time manifold $M$ is considered as a product  of the
time axis $R$ and $n$ manifolds $M_1,\ldots,M_n$, i.e.
\begin{equation} 
M = R \times M_{1} \times \ldots \times M_{n},
\end{equation}
The product of one part of the manifolds gives the external 3-dimensional
space and the remaining part stands for so called internal spaces. The
internal spaces are supposed to be compact. Further, for sake of generality,
we admit that dimensions $N_i={\rm dim}M_i$ for $i=1,\ldots,n$ are arbitrary.

The manifold $M$ is equipped with the metric
\begin{equation} 
g=-{\rm e}^{2\gamma(t)}dt \otimes dt +
\sum_{i=1}^{n}\exp[2{x^{i}}(t)]g^{(i)},
\end{equation}
where $\gamma(t)$ is an arbitrary function determining
the time $t$ and $g^{(i)}$ is the metric on the manifold $M_i$.
We suppose that the manifolds $M_1,\ldots,M_n$
are the Einstein spaces, i.e.
\begin{equation} 
{R_{k_{i}l_{i}}}[g^{(i)}] = \lambda_{i} g^{(i)}_{k_{i}l_{i}}, \ \
k_{i},l_{i}=1,\ldots ,N_{i},\ \ i=1,\ldots,n,
\end{equation}
where $\lambda_i$ is constant. In the special case, when
$M_i$ is a space of
constant Riemann curvature $K_i$ the constant $\lambda_i$ reads:
$\lambda_i= K_i (N_i-1)$ (here $N_i>1$).

Using the assumptions (2.3) we obtain the following
non-zero components of the Ricci tensor for the metric
(2.2) \cite{I92}
\begin{eqnarray} 
R_0^0= {\rm e}^{-2\gamma}\left(
\sum_{i=1}^{n} N_{i}(\dot{x}^{i})^2+  \ddot{\gamma_0}-
\dot{\gamma}\dot{\gamma_0}\right)\\
R_{n_i}^{m_i}=
\left\{
\lambda_{i} \exp[-2x^i] +
\left[\ddot{x}^{i}+ \dot{x}^{i}(\dot{\gamma_0}-\dot{\gamma})
\right]{\rm e}^{-2\gamma}
\right\}
\delta_{n_i}^{m_i}
\end{eqnarray}
where we denoted
\begin{equation} 
\gamma_0=\sum_{i=1}^{n} N_{i}x^{i}.
\end{equation}
Indices $m_i$ and $n_i$ in (2.4),(2.5) for
$i=1,\ldots,n$ run over from ($D-\sum_{j=i}^{n}N_j$) to
($D-\sum_{j=i}^{n}N_j+N_i$) ($D=1+\sum_{i=1}^{n}N_i={\rm dim}M $).

We consider a source of gravitational field in the form of multicomponent
perfect fluid.
The energy-momentum tensor of such source under the comoving
observer condition reads
\begin{eqnarray}
&& T^{M}_{N} = \sum_{\mu =1}^{\bar{m}} T^{M
(\mu)}_{N}, \\
&&\left(T^{M (\mu)}_{N}\right)={\rm diag}\left(-{\rho^{(\mu)}}(t),
 {p_{1}^{(\mu)}}(t) \delta^{k_{1}}_{l_{1}},
\ldots , {p^{(\mu)}_{n}}(t) \delta^{k_{n}}_{l_{n}}\right),
\end{eqnarray}
Furthermore we suppose that for any $\mu$-th component of the perfect
fluid the barotropic equation of state holds
\begin{equation}
{p_{i}^{(\mu)}}(t) = \left(1-{h_{i}^{(\mu)}}\right){\rho^{(\mu)}}(t),\ \
\mu=1,\ldots,\bar{m},
\end{equation}
where
${h_{i}^{(\mu)}} = {\rm const}$. It should be noted that each
$\mu$-th component admits different barotropic equations of state in the
different spaces $M_1,\ldots,M_n$. From the physical viewpoint this
follows from the separation of the internal spaces with respect
to the external one and with respect to each others.

One easily shows that the equation of motion
$\bigtriangledown_{M} T^{M (\mu)}_{0}=0$ for the $\mu$-th component
of the perfect fluid described by the tensor (2.8) reads
\begin{equation}
\dot{\rho}^{(\mu)}+
\sum_{i=1}^{n}N_{i}\dot{x}^{i}(\rho^{(\mu)} + p_{i}^{(\mu)})=0.
\end{equation}
Using the equations of state (2.9), we obtain from (2.10) the following
integrals of motion
\begin{equation}
A^{(\mu)}={\rho^{(\mu)}}
\exp\left[2\gamma_{0} - \sum_{i=1}^{n}N_ih_i^{(\mu)}x^{i}\right]=
{\rm const}.
\end{equation}

The Einstein equations
$R^M_N-R\delta^M_N/2=\kappa^2T^M_N$
($\kappa^2$ is the gravitational constant), can be written as
$R^M_N=\kappa^2[T^M_N-T\delta^M_N/(D-2)]$.
Further we employ the equations
$R^{0}_{0}-R/2=\kappa^{2}T^{0}_{0}$ and
$R^{m_{i}}_{n_{i}}=\kappa^{2}[T^{m_{i}}_{n_{i}}-
T\delta^{m_{i}}_{n_{i}}/(D-2)]$. Using (2.4)-(2.9), we obtain for them
\begin{equation}
\frac{1}{2}\sum_{i,j=1}^{n}G_{ij}\dot{x}^{i}\dot{x}^{j}+ V=0,
\end{equation}
\begin{eqnarray}
\lambda^{i}{\rm e}^{-2x^i}
+[\ddot{x}^{i} + \dot{x}^{i}(\dot{\gamma_0}-\dot{\gamma})]
{\rm e}^{-2\gamma}&=&
- \kappa^{2} \sum_{\mu=1}^{\bar{m}}A^{(\mu)}
\left(h^{(\mu)}_{i} - \frac{\sum_{k=1}^{n}N_{k}h^{(\mu)}_{k}}{D-2}\right)
\nonumber\\
&\times &
\exp\left[\sum_{i=1}^{n}N_ih_i^{(\mu)} x^{i}-2\gamma_{0}\right].
\end{eqnarray}
Here
\begin{equation}
G_{ij}=N_{i}\delta_{ij}-N_{i}N_{j}
\end{equation}
are the components of the minisuperspace metric,
\begin{equation}
V={\rm e}^{2\gamma}\left(
-\frac{1}{2}\sum_{i=1}^{n}\lambda^{i}N_{i}{\rm e}^{-2x^i}+
\kappa^2\sum_{\mu=1}^{\bar{m}}A^{(\mu)}
\exp\left[\sum_{i=1}^{n}N_ih_i^{(\mu)} x^{i}-2\gamma_{0}\right]
\right).
\end{equation}
The dependence on the densities $\rho^{(\mu)}$ in (2.12),(2.13) has been
canceled according to the relations (2.11).

It is not difficult to verify that after the gauge fixing
$\gamma=F(x^1, \ldots, x^n)$ the equations of motion (2.13) may be considered
as the Lagrange-Euler equations obtained from the Lagrangian
\begin{equation}
L={\rm e}^{\gamma_0-\gamma}\left(
\frac{1}{2}\sum_{i,j=1}^{n}G_{ij}\dot{x}^{i}\dot{x}^{j}-V
\right)
\end{equation}
under the zero-energy constraint (2.12).

Now we introduce n-dimensional real vector space $R^{n}$.
By $e_{1},\ldots e_{n}$ we denote the canonical basis in  $R^{n}$
($e_{1}=(1,0,\ldots,0$) etc.). Hereafter we use the following vectors:

\noindent
the vector we need to obtain
\begin{equation}
x=x^1(t)e_1 + \ldots + x^n(t)e_n,
\end{equation}

\noindent
the vector induced by the curvature of the space $M_k$
\begin{equation}
v_k=-\frac{2}{N_k}e_k=\sum_{i=1}^n\frac{-2}{N_k}\delta^i_k e_i,
\end{equation}

\noindent
the vector induced by $\mu$-th component of the perfect fluid
\begin{equation}
u_{\mu}=\sum_{i=1}^n\left(
h^{(\mu)}_{i} - \frac{\sum_{k=1}^{n}N_{k}h^{(\mu)}_{k}}{D-2}
\right)e_i.
\end{equation}
Let $<.,.>$ be a symmetrical bilinear form defined on $R^n$
such that
\begin{equation}
<e_{i},e_{j}>=G_{ij}.
\end{equation}
The form is nongenerated and the inverse matrix to
$(G_{ij})$ has the components
\begin{equation}
G^{ij} = \frac{\delta^{ij}}{N_{i}}+\frac{1}{2-D}.
\end{equation}
The form  $<.,.>$ endows the space $R^n$ with the metric, which signature is
$(-, +, ..., +)$ \cite {I89},\cite {I89a}. By the usual way we may
introduce the covariant components of vectors. For the vectors $v_k$ and
$u_{\mu}$ we have
\begin{eqnarray}
v_{(k)}^i=-2\frac{\delta^i_k}{N_k},\ \
v_i^{(k)}=\sum_{i=1}^n G_{ij}v_{(k)}^j=2(N_i-\delta^k_i),\\
u_{(\mu)}^i=
h^{(\mu)}_{i} - \frac{\sum_{k=1}^{n}N_{k}h^{(\mu)}_{k}}{D-2},\ \
u^{(\mu)}_i=\sum_{i=1}^n G_{ij}u_{(\mu)}^j=N_ih^{(\mu)}_{i}.
\end{eqnarray}
The values of $<v_k,v_i>$, $<v_k,u_{\mu}>$ and $<u_{\mu},u_{\nu}>$ are
presented in Table 1.

\begin{center}
\begin{tabular}{|c|c|c|}
\hline
\hline
& &\\
$<.,.>$ &$v_{j}$&$u_{\nu}$\\
& &\\
\hline
& &\\
$v_{i}$&$4(\frac{\delta_{ij}}{N_{i}}-1)$&$-2h_{i}^{(\nu)}$ \\
& &\\
\hline
& &\\
$u_{\mu}$&$-2h_{j}^{(\mu)}$&$\sum_{i=1}^{n}h_{i}^{(\mu)}
h_{i}^{(\nu)} N_{i}+
\frac{1}{2-D}[\sum_{i=1}^{n}h_{i}^{(\mu)} N_{i}]
[\sum_{j=1}^{n}h_{j}^{(\nu)} N_{j}]$\\
& &\\
\hline
\hline
\end{tabular}
\end{center}

TABLE I. Values of the bilinear form $<.,.>$ for the vectors
$v_i$ and $u_{\mu}$, induced by curvature of the space $M_i$ and
$\mu$-th component of the perfect fluid correspondingly.

A vector $y\in R^{n}$ is called time-like, space-like
or isotropic, if $<y,y>$  has  negative,  positive  or  null values
correspondingly. Vectors $y$ and $z$ are called orthogonal if
$<y,z>=0$. It should be noted that the curvature induced vector
$v_i$ is always time-like, while the perfect fluid induced vector
$u_{\mu}$ admits any value of $<u_{\mu},u_{\mu}>$ (see Table 1).

Using the notation $<.,.>$ and the vectors (2.17)-(2.19), we may write the
zero-energy constraint (2.12) and the Lagrangian (2.16) in the form
\begin{eqnarray}
E=\frac{1}{2}<\dot{x},\dot{x}>+V=0,\\
L={\rm e}^{\gamma_0-\gamma}\left(
\frac{1}{2}<\dot{x},\dot{x}>-V
\right),
\end{eqnarray}
where
\begin{equation}
V={\rm e}^{2(\gamma-\gamma_0)}\left[
-\frac{1}{2}\sum_{i=1}^{n}\lambda^iN_i{\rm e}^{<v_i,x>}+
\kappa^2\sum_{\mu=1}^{\bar{m}}A^{(\mu)}
{\rm e}^{<u_{\mu}, x>}
\right].
\end{equation}
It is obviously from (2.26) that the term induced in the potential by the
non-Ricci flat space $M_i$ is similar to the term induced by $\mu$-component
of the perfect fluid. Due to this fact the non-zero curvature of the
manifold $M_i$ may be also called a component and now we use the notion of
the component in such new sense.  Further we employ the so called harmonic
time gauge, which implies
\begin{equation} 
\gamma(t)=\gamma_0=\sum_{i=1}^{n} N_{i}x^{i}.
\end{equation}

>From the mathematical viewpoint the problem consist in integrability of the
system with $n\geq 2$ degrees of freedom, described by the Lagrangian of the
form
\begin{equation}
L=
\frac{1}{2}<\dot{x},\dot{x}>- \sum_{\mu=1}^ma^{(\mu)}
{\rm e}^{<b_{\mu}, x>},
\end{equation}
where $x,b_{\mu}\in R^n$.
In (2.28) $m$ denotes the total number of the components including the
curvatures and the perfect fluid components.
It should be noted that the kinetic term
$<\dot{x},\dot{x}>$ is not positively definite bilinear form as it usually
takes place in classical mechanics. Due to the pseudo-Euclidean signature
$(-, +, ..., +)$ of the form $<.,.>$ such systems may be called
pseudo-Euclidean Toda-like systems as the potential like that given in
(2.28) defines well known in classical mechanics Toda lattices \cite{Toda}.
In the papers \cite{G95},\cite{I92},\cite{I94}
the following classes of the
integrable pseudo-Euclidean Toda-like systems have been found
\begin{enumerate}
\item
$m=0$. This case corresponds to the vacuum multidimensional cosmological
model on the manifold
$M = R \times M_{1} \times \ldots \times M_{n}$  with all Ricci-flat
spaces $M_{i}$. The corresponding metric is a multidimensional
generalization of the well-known Kasner solution \cite{I92}.

\item
m=1, the vector $b_1$ is arbitrary. The metrics for this 1-component
case were obtained in \cite{I94}. This integrable class may be enlarged by
the addition of the new components inducing the vectors collinear to the
vector $b_1$.

\item
$m\geq 2,\ n=2,\ b_{\mu}=b+C_{\mu}b_0$, where $b$ is an arbitrary vector and
$b_0$ is an arbitrary isotropic vector, $C_{\mu}$=const. This class was
integrated in \cite{I94} only under the zero energy constraint.

\item
$m\geq 2$, the vectors $b_1,\ldots,b_m$ are linear independent and satisfy
the conditions \linebreak
$<b_{\mu},b_{\nu}>=0$ for $\mu\neq\nu$. This integrable class
may be enlarged by the addition of the new components inducing the vectors
collinear to one from the orthogonal set  $b_1,\ldots,b_m$. The
corresponding cosmological models are studied in \cite{G95}.

\item
$m\geq 2$, the vectors $b_1,\ldots,b_m$ are space-like and
may be interpreted as a set
of admissible roots \cite{Bo} of a simple complex Lie's algebra $G$. In this
case the pseudo-Euclidean Toda-like system is trivially reducible to the
Toda lattice associated with the Lie algebra $G$ \cite{Toda}.
For $G=A_2\equiv sl(3,C)$
the metric of the corresponding cosmological model was explicitly written in
\cite{G95}.
\end{enumerate}

In the present paper we consider only 2-component ($m=2$) pseudo-Euclidean
Toda-like systems with 2 degrees of freedom ($n=2$) under the zero energy
constraint. The corresponding multidimensional cosmological models are
2-factor spaces, i.e.
\begin{equation}
M = R \times M_{1} \times M_{2}
\end{equation}
and admit the following combinations of the components: curvature of $M_1$
and curvature of $M_2$ (vacuum models); curvature of $M_1$ or $M_2$ and
1-component perfect fluid; 2-component perfect fluid in Ricci-flat spaces
$M_1$ and $M_2$. In our recent paper \cite{G96} we have integrated the
vacuum model of the type (2.29) with 2 curvatures for the dimensions
(dim$M_1$,dim$M_2$)=(6,3),(8,2),(5,5). Now we develop more general procedure
useful for any combination of the 2 components.

\section{Reducing to the generalized Emden-Fowler equation}

Let us consider the equations of motion following from the Lagrangian (2.28)
with $n=m=2$ under the zero energy constraint. If the vectors $b_1$ and
$b_2$ satisfy one of the following conditions
\begin{enumerate}
\item
$b_1$ and $b_2$ are linearly dependent,
\item
$<b_1,b_2>=0$, i.e. $b_1$ and $b_2$ are orthogonal ,
\item
$<b_1-b_2,b_1-b_2>=0$, i.e vector $b_1-b_2$ is isotropic,
\end{enumerate}
the equations of motion are easily integrable and the corresponding exact
solutions have been obtained in the papers \cite{G95},\cite{I94}. Now we
aim to develop the integration procedure just for all remaining cases. Then,
further we suppose that the vectors $b_1$ and $b_2$ do not satisfy any
condition from 1-3.

Let us introduce in $R^2$ an orthogonal basis forming by the following two
vectors
\begin{equation}
f_1=(u_{22}-u_{12})b_1 + (u_{11}-u_{12})b_2,\ \
f_2=b_2-b_1,
\end{equation}
where we denoted
\begin{equation}
u_{\mu\nu}=<b_{\mu},b_{\nu}>, \ \ \mu,\nu=1,2.
\end{equation}
According to the admission accepted $f_2$ is not isotropic vector, i.e.
\begin{equation}
<f_2,f_2>=u_{11}+u_{22}-2u_{12}\neq 0.
\end{equation}
One may easily check that $u_{12}^2-u_{11}u_{22}\geq 0$ for any vectors
$b_1,b_2\in R^2$ and $u_{12}^2-u_{11}u_{22}=0$ if and only if $b_1$ and
$b_2$ are linearly dependent. Then in the case under consideration
\begin{eqnarray}
<f_1,f_1>=-(u_{12}^2-u_{11}u_{22})(u_{11}+u_{22}-2u_{12})\neq 0,\\
<f_1,f_1>/<f_2,f_2>=-(u_{12}^2-u_{11}u_{22})< 0,
\end{eqnarray}
i.e. one from the orthogonal vectors $f_1$ and $f_2$ is space-like and the
other is time-like.

The vector $x(t)$ we have to find decomposes as follows
\begin{equation}
x=\frac{<x,f_1>}{<f_1,f_1>}f_1 + \frac{<x,f_2>}{<f_2,f_2>}f_2.
\end{equation}
For the new configuration variables
\begin{eqnarray}
z(t)=\frac{<x,f_2>}{2}+\ln{\sqrt{\left|\frac{a^{(2)}}{a^{(1)}}\right|}},\\
y(t)=\frac{1}{2}\sqrt{-\frac{<f_2,f_2>}{<f_1,f_1>}}<x,f_1>
\end{eqnarray}
the Lagrangian (2.28) and the corresponding zero-energy constraint look as
follows
\begin{eqnarray}
L=2\beta\left(\dot{z}^2-\dot{y}^2\right)-V(z,y),\\
E=2\beta\left(\dot{z}^2-\dot{y}^2\right)+V(z,y)=0,
\end{eqnarray}
where the potential $V(z,y)$ has the form
\begin{equation}
V(z,y)=V_0{\rm e}^{2\alpha\beta y}
\left(
{\rm sgn}\left[a^{(1)}\right]{\rm e}^{2\beta_1 z} +
{\rm sgn}\left[a^{(2)}\right]{\rm e}^{2\beta_2 z}
\right).
\end{equation}
In formulas (3.9)-(3.11) the following constants are used
\begin{eqnarray}
&\alpha=\sqrt{u_{12}^2-u_{11}u_{22}},
\ \ &\beta=(u_{11}+u_{22}-2u_{12})^{-1},\\
&\beta_1=-(u_{11}-u_{12})\beta,\ \
&\beta_2=\beta_1+1=(u_{22}-u_{12})\beta,\\
&V_0=|a^{(1)}|^{\beta_2}|a^{(2)}|^{-\beta_1}.&
\end{eqnarray}
It should be mentioned that using of a basis in the form (3.1) provides the
factorization of the potential (3.11) with respect to the coordinates of the
vector $x(t)$ (the additional linear transformation (3.7),(3.8) does not
matter in this situation). Such factorization of the potential is essential
under the developing of the following procedure proposed in \cite{Bl91}.
Using the equation of motion following from the Lagrangian (3.9)
\begin{eqnarray}
\ddot{z}&=&-\frac{1}{2\beta}
V_0{\rm e}^{2\alpha\beta y}
\left(
\beta_1{\rm sgn}\left[a^{(1)}\right]{\rm e}^{2\beta_1 z} +
\beta_2{\rm sgn}\left[a^{(2)}\right]{\rm e}^{2\beta_2 z}
\right),\\
\ddot{y}&=&\frac{\alpha}{4}V(z,y),
\end{eqnarray}
the zero-energy condition (3.10) written in the form
\begin{equation}
\dot{z}^2=\frac{1}{2\beta}\frac{V(z,y)}{(\dot{y}/\dot{z})^2-1}=
\frac{1}{2\beta}\frac{V(z,y)}{({\rm d}y/{\rm d}z)^2-1}
\end{equation}
and the relation
\begin{equation}
\frac{{\rm d}^2y}{{\rm d}z^2}=\frac{\ddot{y}-\ddot{z}
\frac{{\rm d}y}{{\rm d}z}}
{\dot{z}^2}
\end{equation}
we obtain the following second-order ordinary differential equation
\begin{equation}
\frac{{\rm d}^2y}{{\rm d}z^2}=
\left[\left( \frac{{\rm d}y}{{\rm d}z}\right)^2-1\right]
\left\{
\frac{1}{2}\left(\beta_1+\beta_2+
\frac{{\rm e}^{2z}-\varepsilon}{{\rm e}^{2z}+\varepsilon}\right)
\frac{{\rm d}y}{{\rm d}z}+\alpha\beta
\right\},
\end{equation}
where
\begin{eqnarray}
\varepsilon={\rm sgn}\left[a^{(1)}a^{(2)}\right].
\end{eqnarray}
We notice that due to the factorization of the potential the right side of
the equation (3.19) does not contain $y$, so, in fact, the equation is
the first-order one with respect to  ${\rm d}y/{\rm d}z$.

This procedure is valid for the solutions such that $\dot{z}\not\equiv 0$. Under
the zero energy constraint the solutions of (3.15),(3.16) with
$\dot{z}\equiv 0$ gives the following vector $x(t)$
\begin{equation}
x(t)=p\ln|t-t_0|+q,
\end{equation}
where the constant vectors $p,q\in R^2$ are such that
\begin{eqnarray}
&&p=\frac{2}{\alpha^2}f_1,\\
&&{\rm e}^{<q,b_1>}=\frac{\beta_2}{a^{(1)}\alpha^2\beta}>0,\
{\rm e}^{<q,b_2>}=-\frac{\beta_1}{a^{(2)}\alpha^2\beta}>0.
\end{eqnarray}
We note that the exceptional solution (3.21) exists only if the inequalities
in (3.23) are satisfied.
It should be mentioned, that the set of the equations (3.10),(3.15),(3.16)
does not admit static solutions $\dot{z}=\dot{y}\equiv 0$ due to the
condition (3.3). The solutions with $\dot{z}=\pm\dot{y}$ are also
impossible, so using the relation (3.17) we do not lose any solutions of the
set (3.10),(3.15),(3.16) except, possibly, the solution (3.21).

Let us suppose that one is able to obtain the general solution of the
equation (3.19) in the parametrical form $z=z(\tau)$, $y=y(\tau)$,
where $\tau$ is a parameter. Then using (3.6)-(3.8) we obtain the vector $x$
as the function of the parameter $\tau$
\begin{equation}
x(\tau)=\frac{2y(\tau)}{\alpha}\left(-\beta_2b_1+\beta_1b_2\right)
+2\beta\left[z(\tau)-\ln{\sqrt{\left|\frac{a^{(2)}}{a^{(1)}}\right|}}
\right](b_2-b_1).
\end{equation}
We recall that coordinates of the vector $x(\tau)$ in the canonical
basis are the logarithms of the scale factors for the spaces
$M_1,M_2$.
The relation between the harmonic time $t$ and the parameter $\tau$ may be
always derived by integration of the zero-energy constraint written in the
form of the separable equation
\begin{equation}
{\rm d}t^2=2\beta\frac
{\left(\frac{{\rm d}y}{{\rm d}\tau}\right)^2-
\left(\frac{{\rm d}z}{{\rm d}\tau}\right)^2}
{V(z(\tau),y(\tau))}
{\rm d}\tau^2.
\end{equation}
Thus the problem of the integrability by quadrature of the pseudo-Euclidean
Toda-like systems with 2 degrees of freedom under the zero-energy constraint
is reduced to the integrability of the equation (3.19).

For ${\rm d}y/{\rm d}z$ the equation (3.19) represents the first-order
nonlinear ordinary differential equation. Its right side is third-order
polynom (with the coefficients depending on $z$) with respect to the
${\rm d}y/{\rm d}z$. An equation of such type is called Abel's equation
(see, for instance \cite{PZ},\cite{ZP}).
There are no methods to integrate arbitrary Abel's
equation, however the equation (3.19) may be integrated for some values of
the parameters $\alpha\beta$ and $\beta_1+\beta_2$. First of all let us
notice that the equation (3.19) has the partial integrals $y\pm z=$const,
which make the relation (3.17) singular and as was already mentioned are not
partial integrals of the set (3.10),(3.15),(3.16). Existence of
this partial solution of the Abel equation (3.19) allows to find the
following nontrivial transformation
\begin{eqnarray}
{\rm e}^{2z}&=&-\varepsilon\frac{X}{Y}\frac{{\rm d}Y}{{\rm d}X},\\
y&=&\delta\left[
z+\ln\left|\frac{Y}{X}\right| +\ln C\right], \ \ \delta=\pm 1,\ \ C>0,
\end{eqnarray}
which reduces the Abel equation (3.19) to the so called generalized
Emden-Fowler equation
\begin{equation}
\frac{{\rm d}^2Y}{{\rm d}X^2}=X^nY^m\left(\frac{{\rm d}Y}{{\rm
d}X}\right)^l,
\end{equation}
where the constant parameters $n,m$ and $l$ read
\begin{eqnarray}
n&=&\frac{1}{2}\left(\beta_1+\beta_2-2\delta\alpha\beta-3\right)=
\frac
{ -2u_{11}-u_{22}+3u_{12}-\delta\sqrt{u_{12}^2-u_{11}u_{22}} }
{ u_{11}+u_{22}-2u_{12} },\\
m&=-&\frac{1}{2}\left(\beta_1+\beta_2-2\delta\alpha\beta+3\right)=
\frac
{ -u_{11}-2u_{22}+3u_{12}+\delta\sqrt{u_{12}^2-u_{11}u_{22}} }
{ u_{11}+u_{22}-2u_{12} },\\
l&=-&\frac{1}{2}\left(\beta_1+\beta_2+2\delta\alpha\beta-3\right)=
\frac
{ 2u_{11}+u_{22}-3u_{12}-\delta\sqrt{u_{12}^2-u_{11}u_{22}} }
{ u_{11}+u_{22}-2u_{12} }.
\end{eqnarray}
For our models the parameters in the generalized Emden-Fawler equation are
not independent. It follows from (3.29),(3.30) that
\begin{equation}
n+m=-3.
\end{equation}
In the special case $l=0$ the equation (3.28) is known as the Emden-Fowler
equation.

If the parameters $l$ and $m$ given by (3.31),(3.30)
are such that $l=0,\ m\neq 1$ there exists
one more transformation
\begin{eqnarray}
&&1+\varepsilon{\rm e}^{2z}=
-\frac{2}{m-1}\frac{X}{Y}\frac{{\rm d}Y}{{\rm d}X},\\
&&y=\delta\left[ z-\frac{1}{m-1}\ln Y^2 +C\right], \
\delta=\pm 1,\ C\in R,
\end{eqnarray}
which reduces the Abel equation (3.19) to the following
integrable Emden-Fowler equation
\begin{equation}
\frac{{\rm d}^2Y}{{\rm
d}X^2}=Y^{\frac{m+3}{m-1}}.
\end{equation}
There are no methods for
integrating of the generalized Emden-Fowler equation with arbitrary
independent parameters $n,m$ and $l$. However, the discrete-group  methods
recently developed by Zaitsev and Polyanin \cite{ZP} allows to integrate by
quadrature 3 two-parametrical classes, 11 one-parametrical classes and about
90 separated points in the parametrical space $(n,m,l)$ of the generalized
Emden-Fowler equation. For instance, the two-parametrical integrable classes
arise when $m$ and $l$ are arbitrary and $n=0$ or when $n$ and $l$ are
arbitrary and $m=0$. The one-parametrical class with $l=0$ and $n+m=-3$ is
also integrable by quadrature.

Let us suppose that the two components of the 2-factor spaces cosmological
model under consideration induce such vectors $b_1$ and $b_2$ that the
corresponding to the model generalized Emden-Fowler equation (3.28) with the
parameters defined by (3.2),(3.29)-(3.31) is integrable in the parametrical
form $X=X(\tau)$, $Y=Y(\tau)$, where $\tau$ is a parameter. Then, using the
parameter $\tau$ as the new time coordinate we obtain by the formulas
(3.26),(3.27),(3.24),(3.25) the following final result for the metric (2.2)
\begin{equation} 
g=-f^2(\tau)[a_1(\tau)]^{2N_1}[a_2(\tau)]^{2N_2}d\tau \otimes d\tau +
[a_1(\tau)]^2g^{(1)} + [a_2(\tau)]^2g^{(2)},
\end{equation}
where we denoted
\begin{eqnarray} 
&&f(\tau)=\sqrt{\frac{2|\beta|}{V_0}C^{n+l}}
\left[\frac{Y^{\prime}(\tau)}{X^{\prime}(\tau)}\right]^l
\frac{\left[{X^{\prime}(\tau)}\right]^2}
{X(\tau)Y(\tau)Y^{\prime}(\tau)},\\
&&a_i(\tau)\equiv{\rm e}^{x^i(\tau)}={\rm e}^{\gamma^i}
\left\{
\left|\frac{Y^{\prime}(\tau)}{X^{\prime}(\tau)}\right|^
{ (2-l)b_{(1)}^i-(1-l)b_{(2)}^i }
\left|\frac{Y(\tau)}{X(\tau)}\right|^
{ (2+n)b_{(1)}^i-(1+n)b_{(2)}^i }
\right\}^
\frac{2\beta}{l+n}.
\end{eqnarray}
By $\gamma^i$ for $i=1,2$ we denoted the following constants
\begin{equation} 
\gamma^i=2\beta\left\{
\frac{\ln C}{n+l}
\left[ (n-l+4)b_{(1)}^i-(n-l+2)b_{(2)}^i \right] -
\ln{\sqrt{\left|\frac{a^{(2)}}{a^{(1)}}\right|}}
\left[ b_{(2)}^i-b_{(1)}^i \right]
\right\}.
\end{equation}
We recall that $b^i_{(\mu)}$ are coordinates of the vector $b_{\mu}$
in the canonical basis.
In the special case $l=0$ one may also use by the similar manner the
transformation (3.33),(3.34) and the result of integration of the equation
(3.35) to write the metric. This transformation was used in \cite{G96}
for integrating of the models with two curvatures.

Thus the method described allows to integrate the cosmological models if the
corresponding generalized Emden-Fowler equation is integrable. Note that if
the model with some vectors $b_1$ and $b_2$ is integrable by such manner
then any model with the vectors $\alpha b_1$ and $\alpha b_2$ ($\alpha$ is
an arbitrary non-zero constant) is also integrable as the parameters $n,m$
and $l$ do not change under such transformation of the vectors. Taking into
account the classes 1-4 (the class 5 does not arise for $n=2$) and the
additional to them class, which may be integrated by the method described,
we obtain the quite large variety of the integrable 2-factor spaces
cosmological models with 2 components.
\section{Examples of the integrable models}
Now we apply the method proposed in Section 3 to the cosmological models on
the manifold (2.29) with both Ricci-flat spaces $M_1,M_2$ and the
2-component perfect fluid source. Let us represent such model by Table 2

\begin{center}
\begin{tabular}{|c|c|c|} \hline & &\\
{\large manifold/source}&
external space $M_1^{N_1}$&internal space $M_2^{N_2}$\\
& &\\
\hline
\hline
& &\\
1-st component of& &\\
the perfect fluid&$h_1^{(1)}$&$h_2^{(1)}$\\
& &\\
\hline
& &\\
2-nd component of& &\\
the perfect fluid&$h_1^{(2)}$&$h_2^{(2)}$\\
& &\\
\hline
\end{tabular}
\end{center}

TABLE 2. Representation of the model on the manifold
$M = R \times M_{1} \times M_{2}$ with Ricci-flat spaces $M_1,M_2$
for the 2-component perfect fluid.

We recall that $N_i=$dim$M_i$ and $h_i^{(\mu)}$ are the constant parameters
in the barotropic equation of state (2.9). The model is entirely defined by
these 6 parameters. One easily shows \cite{G96b} that the dominant energy
condition applied to the stress-energy tensor (2.8) implies
$0\leq h_i^{(\mu)}\leq 2$. Usually rational values of the parameter
$h_i^{(\mu)}$ are employed in cosmology, for instance,
$h_i^{(\mu)}=(N_i-1)/N_i$ - radiation, $h_i^{(\mu)}=1$ - dust,
$h_i^{(\mu)}=0$ - Zeldovich (stiff) matter, $h_i^{(\mu)}=2$ - false
vacuum ($\Lambda$-term), $h_i^{(\mu)}=(D-1)/D$ - superradiation etc.
On the other hand the most known cases, when the generalized Emden-Fowler
equation (3.28) is integrable, arise for the rational parameters
$n,m$ and $l$. So if one demands the rationality of the parameters $n,m$ and
$l$ in the equation (3.28) corresponding to the model under the condition of
the rationality for the parameters $h_i^{(\mu)}$, then due to the following
relation
\begin{equation} 
\alpha^2=u_{12}^2-u_{11}u_{22}=\frac{N_1N_2}{N_1+N_2-1}
\left(
h_1^{(1)}h_2^{(2)}-h_2^{(1)}h_1^{(2)}
\right)^2,
\end{equation}
the dimensions $N_1,N_2$ with integer value of the expression
$R\equiv\sqrt{N_1N_2(N_1+N_2-1)}$ are singled out. For instance, the
expression $R$ is integer for the following dimensions:
$(N_1,N_2)=(3,6),(2,8),(5,5),(7,8),(3,25),(N_1,1)$. From the physical
viewpoint the following cases may be of interest:
$(2,1),(3,1),(3,6),(3,25)$.

Let us consider the models of the type represented  in Table 2 leading to the
generalized Emden-Fowler equation (3.28) with
\begin{equation} 
l=0.
\end{equation}
Due to the relation (3.32) arising for our models the equation is integrable
for arbitrary parameter $m$ and its exact solution has been written by
Zaitsev and Polyanin \cite{PZ}. It is worth to mention that other 2
integrable classes of the generalized Emden-Fowler equation (3.28), arising
when $n=0$ or $m=0$, describe the same cosmological models. It follows
easily from (3.29),(3.31) that if the model is such that $l=0$ for
$\delta=1$ (or $\delta=-1$) then $n=0$ for $\delta=-1$ (correspondingly,
$\delta=1$). It is easy to see also from (3.30),(3.31) that the condition
$l=0$ transforms to the condition $m=0$ under the inverse numbering of the
components. Thus, from 3 integrable classes, arising for $n=0$, $m=0$ and
$l=0$, correspondingly, of the generalized Emden-Fowler equation (3.28)
for our models
it is enough to study any one from them, let it be the class with $l=0$. In
this case the equation (3.28) has the form
\begin{equation}
\frac{{\rm
d}^2Y}{{\rm d}X^2}=X^{-m-3}Y^m, \ \ m=-\beta_1-\beta_2.
\end{equation}
By the following transformation
\begin{equation} Y=\frac{\tau}{\xi},\ \
X=\frac{1}{\xi}
\end{equation}
it reduces to the equation
\begin{equation}
\frac{{\rm d}^2\tau}{{\rm d}\xi^2}=\tau^m,
\end{equation}
which is easily integrable. Then the general solution of the equation (4.3)
has the form
\begin{equation}
Y=\pm\frac{\tau}{F(\tau)},\ \ X=\pm\frac{1}{F(\tau)},
\end{equation}
where
\begin{eqnarray} 
F(\tau)&=&\int\left[\frac{2}{m+1}\tau^{m+1}+C_1\right]^{-1/2}
{\rm d}\tau + C_2, \ \ m\neq-1,\\
&=&\int\left[2\ln|\tau|+C_1\right]^{-1/2}
{\rm d}\tau + C_2, \ \ m=-1.
\end{eqnarray}
We suppose that the both components of the perfect fluid have the positive
mass-energy densities given by (2.11). It means
$a^{(\mu)}=\kappa^2A^{(\mu)}>0$ for $\mu=1,2$, so
$\varepsilon={\rm sgn}\left[a^{(1)}a^{(2)}\right]=1$. Then taking into
account the formula (3.26), we must consider the general solution (4.6) on
such interval of the variable $\tau$ where
\begin{equation}
G(\tau)\equiv -\frac{X(\tau)Y^{\prime}(\tau)}
{Y(\tau)X^{\prime}(\tau)}=
\frac{F(\tau)}{\tau F^{\prime}(\tau)}-1>0.
\end{equation}

Finally using the results of Section 3 we obtain the following exact
solution for the cosmological model represented by Table 2 in the special
case $l=0$:

\noindent
the metric is given by the formula (3.36), where
\begin{equation}
f^2(\tau)=\frac{2|\beta|C^{-2\delta\alpha\beta}}
{\left[A^{(1)}\right]^{\beta_2}\left[A^{(2)}\right]^{-\beta_1}}
\left[\frac{F^{\prime}(\tau)}{G(\tau)\tau^2}\right]^2,
\end{equation}

\noindent
the scale factors of the spaces $M_1,M_2$ have the form
\begin{equation}
a_i(\tau)=
\left[\frac{A^{(1)}}{A^{(2)}}\right]^{\beta\left[u_{(2)}^i-u_{(1)}^i\right]}
\left\{
\left[G(\tau)\right]^{u_{(2)}^i-2u_{(1)}^i}
\left[C^2\tau^2\right]^{\beta_1u_{(2)}^i-\beta_2u_{(1)}^i}
\right\}^{\delta /\alpha},
\end{equation}

\noindent
the mass-energy densities of the components read
\begin{eqnarray}
\rho^{(1)}(\tau)=
\frac{\left[A^{(1)}\right]^{\beta_2}\left[A^{(2)}\right]^{-\beta_1}}
{[a_1(\tau)]^{2N_1}[a_2(\tau)]^{2N_2}}
\left\{
\left[G(\tau)\right]^{u_{12}-2u_{11}}
\left[C^2\tau^2\right]^{\alpha^2\beta}
\right\}^{\delta/\alpha},\\
\rho^{(2)}(\tau)=
\frac{\left[A^{(1)}\right]^{\beta_2}\left[A^{(2)}\right]^{-\beta_1}}
{[a_1(\tau)]^{2N_1}[a_2(\tau)]^{2N_2}}
\left\{
\left[G(\tau)\right]^{u_{22}-2u_{12}}
\left[C^2\tau^2\right]^{\alpha^2\beta}
\right\}^{\delta/\alpha}.
\end{eqnarray}
The functions $F(\tau)$ and $G(\tau)$ are defined in (4.7)-(4.9); components
$u^i_{(\mu)}$ of the vectors in the canonical basis are given in (2.23);
the parameters $\alpha,\beta,\beta_1,\beta_2$ are defined in (3.12),(3.13);
the values $u_{\mu\nu}=<u_{\mu},u_{\nu}>$ for $\mu,\nu=1,2$ may be
calculated from the parameters $N_i,h_i^{(\mu)}$ by the formula given in
Table 1.

The following relation is valid for the densities
(4.12)-(4.13)
\begin{equation}
\rho^{(2)}(\tau)/\rho^{(1)}(\tau)=G(\tau).
\end{equation}
We recall the possible existence of the special solution (3.21).

Concluding the paper, we mention in Table 3 some interesting from our
viewpoint special models with $l=0$ for the dimensions $N_1=3$ and
$N_2=6$. One easily shows that for these dimensions and given in Table 3
values of the parameters $h_i^{(\mu)}$ the parameter $l$  given by (3.31) is
equal to zero, so the special model is described by the
exact solution (4.10)-(4.13). It follows from (2.22),(2.23) that for these
dimensions the 2 perfect fluid components with the parameters 1-st:
$h_1^{(1)}=4/3,h_2^{(1)}=2$, 2-nd: $h_1^{(2)}=2,h_2^{(1)}=5/3$ induce the
vectors, which coincide with the vectors induced by the curvatures of
$M_1^3$ and $M_2^3$, correspondingly. Then the adding of such 2 components
to the integrable vacuum model on $R\times M_1^3\times M_2^6$ with 2
curvatures (see investigation of this model in \cite{G96}) provides with the
integrable model for the 2 non Ricci-flat spaces and the 2-component perfect
fluid.

\begin{center}
\begin{tabular}{|c|c|c|}
\hline
& &\\
{\large manifold/source}&
external space $M_1^3$&internal space $M_2^6$\\
& &\\
\hline
\hline
1-st component of&radiation &radiation\\
the perfect fluid&$h_1^{(1)}=\frac{2}{3}$&$h_2^{(1)}=\frac{5}{6}$\\
\hline

2-nd component of&radiation &Zeldovich matter\\
the perfect fluid&$h_1^{(2)}=\frac{2}{3}$&$h_2^{(2)}=0$\\
\hline
\hline
1-st component of&dust &radiation\\
the perfect fluid&$h_1^{(1)}=1$&$h_2^{(1)}=\frac{5}{6}$\\
\hline
2-nd component of&radiation &dust\\
the perfect fluid&$h_1^{(2)}=\frac{2}{3}$&$h_2^{(2)}=1$\\
\hline
\hline
1-st component of&radiation &radiation\\
the perfect fluid&$h_1^{(1)}=\frac{2}{3}$&$h_2^{(1)}=\frac{5}{6}$\\
\hline
2-nd component of&dust&radiation\\
the perfect fluid&$h_1^{(2)}=1$&$h_2^{(2)}=\frac{5}{6}$\\
\hline
\hline
1-st component of&radiation &radiation\\
the perfect fluid&$h_1^{(1)}=\frac{2}{3}$&$h_2^{(1)}=\frac{5}{6}$\\
\hline
2-nd component of&false vacuum&false vacuum\\
the perfect fluid&$h_1^{(2)}=2$&$h_2^{(2)}=2$\\
\hline
\hline
1-st component of&false vacuum &false vacuum\\
the perfect fluid&$h_1^{(1)}=2$&$h_2^{(1)}=2$\\
\hline
2-nd component of&Zeldovich matter&false vacuum\\
the perfect fluid&$h_1^{(2)}=0$&$h_2^{(2)}=2$\\
\hline
\end{tabular}
\end{center}

TABLE 3. Examples of the integrable models for dimensions
$N_1=3$ and $N_2=6$. The corresponding exact solutions are given by the
formulas (4.10)-(4.13).

\begin{center}
{\bf Acknowledgments}
\end{center}
\par
The authors are grateful to Dr. M. Rainer for useful remarks.
\par

\pagebreak

\end{document}